\documentstyle[12pt]{article}
\begin{document}

\begin{center}
\LARGE {\bf The Imaginary Sliding Window As a New
Data Structure
for Adaptive Algorithms  }\\[5mm]

\Large  {Boris Ryabko }\\[3mm]
\date{}

\normalsize
Siberian State
University \\ of Telecommunication and Computer Science.
\end{center}

\begin{center}
\begin{minipage}[t]{12cm}\small

{\bf Abstract.}\footnotemark\
The scheme of the sliding window is known in Information Theory,
Computer Science, the problem of predicting and in stastistics. Let a source with
unknown statistics generate some word $\ldots x_{-1}x_{0}x_{1}x_{2}\ldots$
in some alphabet $A$. For every moment $t, t=\ldots $ $-1, 0, 1, \ldots$, one stores
the word ("window") $ x_{t-w} x_{t-w+1}\ldots x_{t-1}$ where $w$,$w \geq 1$, is
called "window length". In the theory of universal coding, the code of the
$x_{t}$ depends on source ststistics estimated by the window, in the problem of
predicting, each letter $x_{t}$ is predicted using information of the window, etc.
After that the letter $x_{t}$ is included in the window on the right, while
$x_{t-w}$ is removed from the window. It is the  sliding window  scheme.
 This  scheme   has  two  merits:  it  allows  one  i)  to
estimate the source statistics quite  precisely   and  ii)  to
adapt  the code  in  case  of  a   change  in   the   source'
statistics. However this scheme has a defect, namely, the
necessity to store the window (i.e.  the  word $x_{t-w}\ldots x_{t-1})$
which needs a large memory  size for large $w$. A  new scheme
named  "the Imaginary Sliding  Window   (ISW)" is  constructed.
The gist of this scheme is that not the last element $x_{t-w}$ but
rather a random one is removed from  the window. This  allows one
to  retain  both  merits  of  the  sliding  window as  well  as
the possibility of not storing the window and thus significantly
decreasing  the  memory  size.


\end{minipage}
\end{center}

\bigskip

\footnotetext
      {Supported by Russian Foundation of Basis
       Research  under Grant 99-01-00586 and INTAS under Grant 00-738}

\section{Introduction}

There are many situations when people deal with information sources
with unknown or changing ststistics. Among them we mention data
compression [2], the problem of predicting [14] and the similar problem
of prefetching [15], the problem of adaptive search [9] as well as a statistical
estimation of parameters of the information sources. There are many interesting
ideas and algorithms for solving these problems. For example,
for  encoding  of  information  sources  with  unknown  or
   changing statistics different methods are  used  to  adapt  a
     code to statistics of a source.  Many  of  such  methods  are
  based on context - tree weighting procedure [10],
   on Lempel - Ziv codes [16], see also the review
 in [2] , a bookstack scheme
[11]\footnotemark, on a scheme of sliding window and some others.

 Some of these methods are used  not  only  for  information
sources  encoding,  but  also  for  solving  problems
connected with storage and  search   of  the  information,  see
[1], [9], [15]
  as  well  as  with   statistical  estimation
of  parameters of changing random  processes.
\footnotetext {The bookstack scheme was proposed in authors' paper [11] and
then rediscovered in 1986-1987 in [3] and [4] (see also [12]). Now
 this scheme is usually named "move-to-front scheme" as proposed in
[3].}

The scheme of sliding window is quite popular and may be used jointly with
the adaptive Huffman code [7], the arithmetic code [2], the interval code [4],
the fast code [13], as well as for predicting [14] and prefetching [15] and
other algorithms.

Let us define the sliding window scheme (SW). A source generates the word\\
$\ldots x_{-1}x_{0}x_{1}x_{2}\ldots $ in a finite alphabet $A$. There is a computer
which stores the window $x_{t-w}x_{t-w+1}\ldots x_{t-1}$ of the moment $t$,
where $w (w\geq 1)$ is the length of the window. The computer uses the window
in order to estimate the source statistics. After that the computer moves the window
as follows: the letter $x_{t}$ is included in the window on the right, while
the letter $x_{t-w}$ is removed from the window. If the SW is used for data
compression there exist two computers (an encoder and a decoder).
The decoder conducts  the same operation with the window and this allows to
decode a message definitely.
Naturally, the greater $w$, the more precise the estimate of statistics of the
source.

 Let us consider as an example the problem  of  encoding
the Bernoulli source generating letters  from  some  alphabet
$A=\{a_{1}, a_{2},\ldots ,a_{m} \}$. In this case, the redundancy per letter  for
the best universal code is  not  less  than   $(m-1)/2w+O(1/w)$
when $w$ is the length of the window (see [8]).
(In case of predicting and prefetching the redundancy is equal to the
precision of the prediction).
Hence, to achieve smaller  redundancy or precision,  the  length  of  the
window has to be much greater than the number of  letters  in
the alphabet $A$. Obviously, to keep the window, the encoder and
the decoder need  $w \log m$  bits of memory.

    By $\nu^{t}(a)$ we denote the frequency of occurence of the letter $a$ in
the window $x_{t-w}\ldots x_{t-1}$ for every $a\in A$ . It is known that  a
vector $\{\nu^{t}(a),a \in A\}$  is  a  sufficient  statistic for  the
Bernoulli source (see [6] for example).
Informally,  this means  that  this  vector  implies that all  the
information is contained in $x_{t-w}\ldots x_{t-1}$.  However  to  keep
frequencies  only   $m\log w$   bits   are  needed (which is
exponentially less then  $w \log m$). In the sliding window
scheme after the encoding of the  recurrent  letter   $x_{t}$   the
relevant frequency  increases by $1$ $(\nu^{t+1}(x_{t})= \nu (x_{t})+1)$
while the frequency of the letter $x_{t-w}$, being  removed  from
the window, decreases by $1$.

     Informally, in  the  ISW  scheme,  we  propose  only  to
keep the vector of frequencies $(\bar{\nu}^{t}(a_{1}),\bar{\nu}^{t}(a_{2}),
\ldots ,\bar{\nu}^{t}(a_{m}))$, $\sum_{i=1}^{m} \bar{\nu}^{t}(a_{i})=w$.
After the coding of the recurrent letter $x_{t}$   we
increase its occurrence  by $1$  (as before, $\bar{\nu}^{t+1}(x_{t})=
\bar{\nu}^{t}(x_{t})+ 1)$ and  then  we  decrease   the  occurrence  of
a randomly selected letter by $1$, where  the probability of
decreasing the occurrence of the letter $a\in A$  is  equal  to
$\bar{\nu}^{t}(a)/ w$.

    Thus, in the Imaginary Sliding  Window  scheme  only  the
vector of integers $(\bar{\nu}^{t}(a_{1}),\ldots ,$ $\bar{\nu}^{t}(a_{m}))$
is kept. After  the
encoding of $x_{t}$  the number $\bar{\nu}^{t}(x_{t})$ increases by $1$  and  one
randomly selected number decreases by $1$.

    It turns out that this scheme possesses properties  which
are  similar  to  the usual  sliding  window  :   first,   the
distribution of the vector $(\bar{\nu}^{t}(a_{1}),\ldots ,\bar
{\nu}^{t}(a_{m}))$ is
similar to the
one  for the sliding window and, second, under the changing
of  the  source  statistics, the  adaptation  of  the   vector
$(\bar{\nu}^{t}(a_{1}),\ldots ,\bar{\nu}^{t}(a_{m}))$ occurs.

    The  construction of the ISW scheme is based  on the  use
of random numbers, or, more precisely,  on  a  sequence  of
random  equi   -   probable   independent   binary    digits.

    There are two usual ways to  obtain  random  digits:  the
first, using a table of random digits,  and,  the  second,
using a generator of random digits.
Both of them may be used for predicting and statistical estimation.
However in case of data compression the sequence of digits has to be the same in
the encoder and the decoder. It is possible to use the third way to obtain
random digits.
In this case the random digits are not ideal,  but  they  may  be
obtained free  "of charge".  The  point  is  that the encoded,
"compressed" sequence is "nearly" random. Moreover, the less
the code  redundancy,   the  nearer   encoded  message is to  a
sequence  of  random  bits   generated  by  tossing   a
symmetric coin. We recommend to use as  random  digits  that
part of the message  being  generated  by  a  source,  which  is
already  encoded.  It  is  important  that this  sequence  is
known  by  the encoder and the decoder, and while coding, it
is  sufficient to keep only a small current  piece of it.

    We turn our attention to the remainder of the paper.
In Section 2 the asymptotic equivalence of the scheme of the
imaginary sliding  window  with  the  usual  scheme  of   the
sliding window is proved and the extention to the  case  of
Markovian sources is given. In Section 3 a  fast  method  to  use
random digits which are necessary for the  Imaginary  Sliding
Windows scheme is  proposed.  Using   this  method  allows the
 processes  of   encoding  and
decoding to procede without delay.

\section {The Scheme of the Imaginary Sliding Window}

     Let us give some necessary definitions.  Denote the  set
of words of the length k in the alphabet $A$ as $A^{k}$ ,$(k\geq 0)$. Let
$\Omega_{\infty}$ be  the  set  of  all  ergodic   and  stationary sources
generating letters from $A$.
   For $\omega \in \Omega_{\infty}$, $a\in A$, $k\geq 0$, $u\in A^{k}$
denote by  $P_{\omega}(a/u)$  the
probability   that letter $a$ is generated  next by the source
$\omega$ in the case when the word $u\in A^{k}$ is generated by the same
source. According to the  definition, $\mu$  will be  the memory
of  the  source  in  the  case  if  for  all  letters  $a$,
$u_{1}, u_{2},\ldots , u_{k}\in A$, $k \geq \mu$ the equality
$$
P_{\omega}(a/u_{1}\ldots u_{k})= P_{\omega}(a/u_{1}\ldots u_{\mu})
$$
is valid ( when $\mu=0$  a source  is  said  to  be  a  Bernoulli
source). Denote  as $\Omega_{\mu}$, $\mu \geq 0$, a set of all
$\omega \in \Omega_{\infty}$   with  the
memory $\mu$.

   Let us  describe the scheme of the ISW for  the  case
of a Bernoulli source. Let $x_{1} x_{2}\ldots x_{t}\ldots$ be the sequence
being generated by some Bernoulli source $\omega \in \Omega_{0}$. Let $w$,
$w\geq1$, be the window length and let for any integer $t$ a  word
$x_{t-w} x_{t-w+1}\ldots x_{t-1}$ be the  window  at  the moment $t$.

   Denote as $\nu_{j}^{t}$ the number of occurences of the letter $a_{j}\in A$
in the window $x_{t-w}\ldots x_{t-1}$. It is easy to see that $(\nu_{1}^{t},
\ldots , \nu_{m}^{t})$
is a  random vector governed by the  multinomial distribution :
\begin{equation}
P \{\nu_{1}^{t}= n_{1},\nu_{2}^{t}=n_{2},\ldots , \nu_{m}^{t}=n_{m}\}=
\left (\begin{array}{cc}
w\\
n_{1}n_{2}\ldots n_{m}
\end{array}
\right ) \prod_{i=1}^{m} P(a_{i})^{n_{i}}
\end{equation}

    In the construction of the ISW  only the set of  integers
(without a window)  $\bar{\nu}^{t} = (\bar{\nu}_{1}^{t},\bar{\nu}_{2}^{t},\ldots ,\bar{\nu}_{m}^{t})$
 is stored,  which
changes after encoding of every letter $x_{t}$.
   To describe the rules of changing  the vector $\bar{\nu}^{t}$ let us
denote  the  random value $\varepsilon^{t}$  being the vector  $1, 2,\ldots ,m$
with the  probabilitis $\bar{\nu}_{1}^{t}/w ,\bar{\nu}_{2}^{t}/w,\ldots ,\bar{\nu}_{m}^{t}/w$,
respectively, i.e.
\begin{equation}
      P \{\varepsilon^{t}=i\}=\bar{\nu}_{i}^{t}/w,\quad i= 1,2,\ldots , m
\end{equation}
After encoding  every letter $x_{t}$ of the message  $x_{1} x_{2}\ldots x_{t}$,
first, $\varepsilon^{t}$ is generated and then the conversion from the vector
$\bar{\nu}^{t}$  to the vector $\bar{\nu}^{t+1}$ is conducted :
\begin{equation}
\bar{\nu}_{j}^{t+1}=
\left \{
\begin{array}{lll}
\bar{\nu}_{j}^{t}-1,\quad \mbox{if}\quad j=\varepsilon^{t}\\
\bar{\nu}_{j}^{t}+1,\quad \mbox{if}\quad a_{j}=x_{t}\\
\nu_{j}^{t}, \qquad\mbox{if}\enskip j\neq \varepsilon^{t}\enskip \mbox{and}\enskip a_{j}\neq x_{t}
\enskip\mbox{or}\enskip j=\varepsilon^{t} \enskip\mbox{and}\enskip x_{t}=a_{j}
\end{array}
\right.
\end{equation}

    In other words, firstly, one random chosen coordinate  of
the vector $\bar{\nu}^{t}$ is decreased  by  $1$.  (This  operation  is
analogous to decreasing  a counter which corresponds to $x_{t-w}$ ,
by $1$,  when  the  window  moves  from  $x_{t-w}\ldots x_{t-1}$ to
$x_{t-w+1}\ldots x_{t}$  and, instead of  removing $x_{t-w}$, one  random
chosen letter is "thrown out"  from the window). Second, the
coordinate of the vector $\bar{\nu}^{t}$ corresponding to the letter $x_{t}$
is increased by $1$.

   The  initial  distribution $\bar{\nu}^{0} =(\bar{\nu}_{1}^{0},\ldots ,\bar{\nu}_{m}^{0})$
 may  be arbitrary chosen. For example, $\bar{\nu}_{i}^{0}= w/m$  may be assumed  when
$i=1,\ldots m$ if $w/m$  is integer.

   Now, we investigate the propeties of the  ISW.  First,  we
demonstrate  that  the  distribution   of   the  vector $\bar{\nu}^{t}$
asymptotically complies with (3), i.e., it is the same  as  the
distribution of the frequency of the occurence of letters  in
the scheme of the sliding window.

{\bf Theorem 1.}  Let a Bernoulli source   be  given  which
generates letters  from  the  alphabet  $A= \{a_{1},\ldots ,a_{m}\}$  with
probabilities  $P(a_{1}),\ldots ,P(a_{m})$, and let $n_{1} n_{2},\ldots ,n_{m}$
be any integer nonnegative numbers such as $\sum n_{i}= w$, $w\geq 1$.
Then for the scheme of the ISW with the vector of frequencies
$\bar{\nu} =(\bar{\nu}_{1}^{t},\ldots ,\bar{\nu}_{m}^{t})$ the following
equality is  valid for  any initial vector $\bar{\nu}^{0} =(\nu_{1}^{0},\ldots ,\nu_{m}^{0})$:
$$
\lim_{t\to \infty} P\left\{\bar{\nu}_{1}^{t}=n_{1},\ldots , \bar{\nu}_{m}^{t}=
n_{m}\right\} =\left (
\begin{array}{cc}
w\\
n_{1},n_{2}\ldots , n_{m}
\end{array}
\right )
\prod_{i=1}^{m} P(a_{1})^{n_{i}}
$$
Proofs of all theorems are given in the  appendix.

 Hence, values $(\bar{\nu}_{1}^{t},\bar{\nu}_{2}^{t},\ldots ,\bar{\nu}_{m}^{t})$
may replace values of
frequency  of occurence of letters in an usual sliding window
and then  the ISW may asymptotically replace the SW.

   The rate of convergence of the distribution
$(\bar{\nu}_{1}^{t},\ldots ,\bar{\nu}_{m}^{t})$
to multinomial distribution (1) comes into the question.  This
is of importance, because this rate  effects  the  rate  of
adaptation of the ISW to modifications of  statistics.  (In
fact, we may assume the statistics  change at the  moment
$t = 0$). We mention two conclusions characterizing the rate  of
approximation  of  frequencies  $(\bar{\nu}_{1}^{t},\ldots ,\bar{\nu}_{m}^{t})$
to the limit
distribution, presupposing that the vector
$(\bar{\nu}_{1}^{0},\ldots ,\bar{\nu}_{m}^{0})$ is
chosen arbitrarily. For simplicity sake, let us define
\begin{equation}
P\left\{
\bar{\nu}_{1}^{\infty}=n_{1},\enskip \bar{\nu}_{2}^{\infty}= n_{2},\ldots ,
\bar{\nu}_{m}^{\infty}=n_{m}
\right \}
=\left (
\begin{array}{cc}
w\\
n_{1},n_{2},\ldots ,n_{m}
\end{array}
\right )
\prod_{i=1}^{m} P(a_{i})^{n_{i}}
\end{equation}

(Such a definition is based  on Theorem 1).

   In Information Theory and Statistics there is  well  known
 the Kullback-Leibler Divergence estimating the divergence  of
the two distributions  of  probabilities.  The  next  Theorem
  allows    estimating  the  divergence  of  the   distribution
$(\bar{\nu}_{1}^{t},\ldots ,\bar{\nu}_{m}^{t})$  to (4).

   {\bf Theorem 2.} Suppose a Bernoulli source generating letters from
the finite alphabet $A= \{ a_{1},\ldots ,a_{m}\}$  is given and we  use  the
scheme  of  ISW  with  the  "window  length" $w$.  Let $R^{t}$   be
Kullback-Leibler   Divergence   between   distributions    of
probabilities of the vector of frequencies
$(\bar{\nu}_{1}^{t},\ldots ,\bar{\nu}_{m}^{t})$ and
$(\bar{\nu}_{1}^{\infty},\ldots ,\bar{\nu}_{m}^{\infty})$,
defined by the equation
\begin{equation}
R^{t}=\sum_{(n_{1},\ldots n_{m})} P\left\{
\bar{\nu}_{1}^{\infty}=n_{1},\ldots ,\bar{\nu}_{m}^{\infty}=n_{m}
\right\}
\log \frac
{P\left\{\bar{\nu}_{1}^{\infty}=n_{1},\ldots ,\bar{\nu}_{m}^{\infty}=n_{m}
\right\} }
{P\left\{\bar{\nu}_{1}^{t}=n_{1},\ldots ,\bar{\nu}_{m}^{t}=n_{m}
\right\} }
\end{equation}
Then, under any initial distribution of frequencies
$(\bar{\nu}_{1}^{0},\ldots ,\bar{\nu}_{m}^{0})$ the inequality
\begin{equation}
R^{t}\leq -\log \left (\sum_{k=0}^{w} \left (
\begin{array}{cc}
w\\
k
\end{array}
\right )
(-1)^{k} \left ( 1-\frac{k}{w}\right )^{t}
\right )
\end{equation}
is valid.

   The right part in (6) is rather  cumbersome. For large $t$
and w the following asymptotic estimate is valid:

   {\it Corollary.} Let $t\to\infty$  and  let
\begin{equation}
\lambda= w\enskip e^{-t/w}.
\end{equation}
Then $ R<\lambda + o(\lambda )$.

It readily follows from the  corollary above that $R^{t}$  becomes
small when $t > w\log w$. If, for example, $t = w \log w + b w$
then $R^{t}$  is close to  $e^{-b}$.

    Thus, the ISW  "remembers" the  initial  distribution  of
probabilities during a period which  is  approximately  equal to
$w \log w$. An "usual" SW can "remember" the initial distribution
of probabilities till total renewal  of  its  contents,  i.e.
till $t = w$.

    To  encode a source as well as to use the schemes of SW
(and  ISW)  in many other  applications  the  estimates  of
probabilities $P(a_{1}),\ldots P(a_{m})$ are used and  the values
$\bar{\nu}_{i}^{t}/w$,
$i=1,\ldots ,m$ (or similar ones) are used as these estimates. The
next Theorem allows  estimation of the proximity of $\bar{\nu}_{i}^{t}/w$ to
$P(a_{i}).$

   {\bf Theorem 3.} Under  fulfilment of the  hypothesises  of  the
Theorem 2,
$$
\mid E(\bar{\nu}_{i}^{t}/w)-P(a_{i})\mid < e^{-t/w}\quad \mbox{for}\enskip
i=1,\ldots , m.
$$

It readily follows from  this  that  the  average  value  of
estimates of probabilities of  the  letters  $a\in  A$
obtained by using  ISW,  quite  rapidly  approximate  to  the
proper value of $P(a )$ under increasing  $t$. It is  important
for application of ISW because $t$ may  be  interpreted  as  the
time duration after  modification  of  statistics  (at  the
moment $t=0$).

    Now, we shall apply the scheme of  ISW  to  the  case  of
Markovian sources. Let $\mu\geq 1$ and it is known that $\omega\in\Omega_{\mu}$.
The construction of ISW described above may be  applied  to  this
case in such a way as while encoding  and  decoding, we  store
$\mid A\mid ^{\mu}$  imaginary windows and each of them  corresponds  to  one
word from $A^{\mu}$. Futhermore, in the memory of the encoder and the
decoder one "real" window is kept, consisting of $\mu$ letters,
and the last $\mu$  letters encoded are stored  in  this  window.
For example, let a source generate the message $x_{1}x_{2}\ldots x_{t}\ldots$
Then, before encoding $x_{t}$  there are  letters $x_{t-\mu}\ldots x_{t-1}$
stored in the "real" window. This word  belongs to $A^{\mu}$,
hence, the ISW correspondsing to it  exists  and  the  letter
$x_{t}$  is encoded in   accordance  with  the  information
stored in this window. After encoding of $x_{t}$, the same  mapping
are made with the ISW which corresponds to $x_{t-\mu}\ldots x_{t-1}$, as
in the Bernoulli case described above. (One  randomly  chosen
frequency decreases at $1$, and a  frequency  corresponding  to
$x_{t}$  increases at $1$).

  Let us consider an  example  which  explains  the  described
construction. Let $ A=\{0,1\}$, $\mu =2$ and let $001011$ be the  sequence  being
encoded. The encoder and the decoder keep in  their
memory $2^{2} = 4$  imaginary windows, and each of them consists of
two nonnegative numbers which,  in  sum,  are  equal  to  the
"window length" $w$ ($w$  is  any  positive  number).  The  first
number corresponds to the frequency of occurence of the
letter $0$  in the window,and the second number corresponds  to
the frequency of occurence of the letter  $1$.  The  letter  $x_{3}$
which follows after 00 is coded on the  basis  of  the  window
corresponding to the word 00, the letter $x_{4}$  is coded  on  the
basis of the contents of the window $01$, etc.

     Thus, while encoding a source of memory $\mu$  we  use
the method which is well known in Information Theory :
represent the Markovian source as a population of Bernoulli
sources. Due to this, every  letter  generated   by  a
source,is encoded and  decoded  according  to  the  information
which is stored in the window corresponding to  the  relevant
Bernoulli source.

\section {Fast Algorithm for Transformation of the ISW}

     After  the  coding  of  every  letter   of   a   message
transformations of frequences of ISW are  conducted:  one
frequency  increases  by $1$  and another,  randomly   chosen,
decreases by $1$. In this section a simple and fast algorithm
of realization of random choice is considered.

    Let   any  generator  of  random  bits  generate  the
sequence $z = z_{1} z_{2}\ldots z_{k}$   which consists of  symbols  from  the
alphabet $\{0,1\}$. We do not estimate the complexity of  generating
these  symbols,
and consider only the method
of transformation of the random bits to meanings of the random values
$\varepsilon^{t}$  (see (4)) which are used for random choice  of the frequency
being decreased by $1$.

    Let us give some  definitions  to  start   describing  an
algorithm. For simplicity sake, we  shall  suppose  that  the
window length $w$ and the number of letters of the  alphabet  $m$
may be represented as
\begin{equation}
w=2^{u},\quad m=2^{\mu}
\end{equation}
when $u$  and $\mu$  are integers.   Let   $\nu^{t} =(\nu_{1}^{t},\ldots ,
\nu_{m}^{t})$ be an
integer-valued vector characterizing the imaginary window.For
generating a meaning of a random value $\varepsilon^{t}$ first, $u$ random
bits $z_{1}\ldots z_{u}$  are produced, and let
$$
z= \sum_{j=1}^{u} z_{j} 2^{u-j}
$$
That  shows,  along  with  (8),  that  $z$,  with   the   same
probability,  may  be  equal  to  any  value  from  the  set
$\{ 0,1,\ldots , w-1\}$, i.e.
\begin{equation}
P\{ z=i\} =\left\{
\begin{array} {ll}
1/w, \enskip \mbox {if} \enskip 0\leq i \leq w-1\\
0, \quad \mbox{for other}\enskip i
\end{array}
\right.
\end{equation}
Let us define
\begin{equation}
Q_{1}=0,\quad Q_{j}=\sum_{k=1}^{j-1}\nu_{k}^{t},\quad j=2,\ldots , m+1
\end{equation}
Let us consider the random value $\varepsilon^{t}$ with the meanings $j$,
$1\leq j\leq m$ if two inequalities hold:
\begin{equation}
Q_{j}\leq z < Q_{j+1}
\end{equation}
From this definition follows:
$$
P \{ \varepsilon^{t}=j \} =P \{ Q_{j}\leq z < Q_{j+1} \} =(Q_{j+1}- Q_{j})/ w
=\nu_{j}/w
$$
(Here the second and the third equalities  follow  from  (9)
and (10)).  Hence, we obtain
$$
 P\{ \varepsilon^{t}= j\} = \nu_{j}^{t}/w
$$
which is the same  as  the  definition  (2).  From  this,  it
follows that the given method of  generating  the  random
value $\varepsilon^{t}$  is quite correct, however it is rather complex.  The
point is that after encoding of the recurrent letter $x_{t}$  from
the message two frequencies must be changed (one has
to be increased by $1$, and one has to be  decreased  by  $1$).
After that, in turn, the values $\{ Q_{j} \}$ must  be
calculated. In the case of a large source
alphabet,  calculation  of  the  value $ Q_{j}$
(according to (10)) and searching $j$ (according to  (11))  may
take too much time. More exactly, $ O(m\enskip \log  w)$  operations
over one-bit words when $m\to \infty$ are needed.

    In conclusion of this section the description of  the
algorithm which allows carring out all operations  with  ISW
during the period $O(\log m \enskip \log w)$ for large  $m$  and $w$,is
given.This algorithm is close to  the  fast  letter-by-letter  code
from the author's  paper [13].

   For description of the method let us define:
\begin{equation}
\Sigma_{1,j}^{t}=\nu_{j}^{t},\enskip l=1,2,\ldots , m;\enskip
\Sigma_{k,j}^{t}=\Sigma_{k-1, 2j-1}^{t}+\Sigma_{k-1,2j}^{t},\enskip
k=2, \ldots , \mu ;\enskip j=1,\ldots, m/2^{k}
\end{equation}
These values are stored in the memory of the encoder and the
decoder. When generating the meaning of the random value $\varepsilon^{t}$
according to $z$, instead of (11) we use the following algorithm:
first, let us check the inequality
\begin{equation}
z \leq \Sigma_{\mu ,1}^{t}
\end{equation}
If it is valid, it means  that $1\leq \varepsilon^{t}\leq m/2$,  otherwise
$m/2 +1\leq \varepsilon^{t}\leq m$. Then, if the inequality  (13)  holds,  we
check  the inequality
\begin{equation}
z\leq \Sigma_{\mu-1, 1}^{t}
\end{equation}
Otherwise  we  calculate $z = z-\Sigma_{\mu ,1}^{t}$ and check the following
condition:
\begin{equation}
z \leq \Sigma_{\mu-1, 3}^{t}
\end{equation}
If (14) holds we  evaluate  whether   the
inequalities $1\leq \varepsilon^{t}\leq m/4$  or $(m/4+1)\leq\varepsilon^{t}\leq m/2$
hold.  If
(15) holds, we obtain $(m/2)+1\leq  3m/4$  or
$(3m/4 +1) \leq\varepsilon^{t}\leq m$. Continuing in that way, after $\log m= \mu$  steps
we shall evaluate $\varepsilon^{t}$. Besides, at every  step,
it is necessary to make one  comparison  and,  possibly,  one
subtraction of numbers each of which has  the form of  a
word of the length $u = \log w$ bits. Thus, the  general  number
of operations over singlebit  words is equal to $O(\log m\enskip \log w)$.

   Now, let us describe the "fast" method of conversion  from
$\left\{\Sigma_{i,j}^{t}\right\}$ to $\left\{\Sigma_{i,j}^{t+1}\right\}$.
Let  under  conversion  from  $t$ to $t+1$
any $j$-coordinate  of  the  vector $\nu^{t}$ increases  by $1$   and
k-coordinate  decreases $(j, k \in \{1,2,\ldots ,m\}$, see  (5)). Then
we have to increase and decrease by $1$ one value from the sets
$$
\left\{ \Sigma_{2,i}^{t},\enskip i=\lambda ,\ldots , m/2\right\} ,\enskip
\left\{\Sigma_{3,i}^{t},\enskip i=1,m/4\right\} ,\enskip\ldots ,
\left\{ \Sigma_{\mu ,i}^{t},\enskip i=1,2\right\}
$$
i.e. make $\mu= \log m$ operations of addition of $1$  and $\mu=\log m$
operations of subtraction of $1$. Each  operation  of  addition
and subtraction is made over the numbers of length $u=\log w$,
so the general number of operations over singlebit  words  is
equal to $O(\log m \enskip\log w)$. Thus, when  using  the  fast  method
proposed  the  number  of  operations  after  encoding  of  a
recurrent letter under transformations of ISW,  is  equal  to
$O(\log m\enskip\log w)$.

\begin{center}
\Large {\bf Appendix}
\end{center}
    Proof of  Theorem 1.

    Denote by $S$  a set of vectors of the form of
$S=(S_{1},\ldots ,S_{m})$ such that all $S_{j}$ are positive integers,
and $\Sigma_{i=1}^{m} S_{j}=w$.
Let us consider a Markov chain  $M$, states  of  which  coincide
with elements of $S$ and a matrix of probabilities  of  conversion
is defined by the equality
\begin{equation}
P_{\sigma ,\delta }=\left\{
\begin{array}{lll}
P (a_{i})\sigma_{j}/w \quad\mbox{if}\enskip \sigma_{1}=\delta_{1},\ldots ,
\delta_{i}=\sigma_{i}+1,\delta_{j}=\sigma_{j}-1 \\
\Sigma_{k=1}^{m}P(a_{k})\delta_{k}/w\quad\mbox{if}\enskip \delta_{1}=\sigma_{1} ,
\enskip \delta_{2}=\sigma_{2},\ldots ,\delta_{m}=\sigma_{m}\\
0\quad\mbox{for another}\enskip \sigma , \delta
\end{array}
\right.
\end{equation}
This Markov chain simulates the behaviour of ISW.

   Using a standard technology of Markov chains (  mentioned,
for example, in [5]), it is easy  to  test  the
assumption that limit probabilities for $M$  exist and  are
established  by the equality
$$
\pi_{\sigma}=\left (
\begin{array}{cc}
w\\
\sigma_{1}\ldots \sigma_{m}
\end{array}
\right )
\prod_{i=1}^{m} P^{\sigma_{i}}(a_{i}),\quad \sigma \in S
$$
which proves  Theorem 1.

       Proof of   Theorem 2.

   Let us introduce a new  scheme --- the sliding window  with
random removing of elements  (SWRRE). In  this  scheme  a
sequence of $w$ "boxes" $(w \geq 1)$ is considered. Every "box"  may
contain a letter from the alphabet $A$. As  above,  a  Bernoulli
source  is given,  generating the sequence $x_{1} x_{2}\ldots$ ,$x_{j}\in A$  for
all $j$, and let $P(a )$ be the probability  of  generating   the
letter $a \in A$.

    In the initial moment there are letters  from $A$  in  the
"boxes". At every moment $t=1,2,\ldots $ two operations are made:  a
random value $\nu^{t}$  is produced which is equal to $1,2,\ldots w$  with
the probability $1/w$  each, and the letter from the box number
$\nu^{t}$ is removed. Then a value $\varphi^{t}$ is produced which is equal  to
$1,2,\ldots m$  such as
$$
P\{ \varphi^{t}=k\} =P(a_{k})
$$
and the letter $a$ is located in the box  which became free.

   It is easy to see that the  scheme SWRRE is an exact  but
more detailed model of the scheme ISW. In fact, denote by $\nu_{k}^{t}$  a
random value which is equal to the number of boxes containing
the letter $a_{k}$  at the moment $t$  and  let
$\nu^{t}= (\nu_{1}^{t},\ldots ,\nu_{m}^{t})$. It
follows  from  the   scheme   SWRRE   described    that   the
probabilities of conversion from $\nu^{t}$  to $\nu^{t+1}$ are also  defined
by the equality (16).  Hence, if the initial distribution  is
the same for both schemes ISW and SWRRE  (i.e. $\bar{\nu}^{0}=\nu^{0})$,
then the distribution of probabilities for all other  moments
will be the same: for any $\bar{n} =(n_{1},\ldots ,n_{m})$
\begin{equation}
P\{ \nu^{t}=\bar{n}\} =P\{\bar{\nu}^{t}=n \} .
\end{equation}

  Then let us introduce  a  new  random  value $\varphi^{t}$ which  is
connected  with SWRRE. By definition, $\varphi^{t}$ is  equal  to  the
number of boxes from which letters were not  removed  at  the
moments $1, 2,\ldots , t$.   Note    immediately   that    the
distribution of this value is well known (see, for  example,
[5]), when $\varphi^{t}$  is the number of  empty  boxes
obtained after the random distribution of  $t$  elements  in  $w$
boxes). It is known that
\begin{equation}
P\left\{ \varphi^{t}=0 \right \}= \sum_{k=0}^{w}(-1)^{k}
\left (
\begin{array}{cc}
w\\
k
\end{array}
\right )
\left ( 1 -\frac{k}{w}\right )^{t}
\end{equation}
(see [5]).

    Let in the scheme of SWRRE all letters be replaced  in
all boxes (i.e. $\varphi^{t} = 0$) at time $t$. Then, obviously, the
distribution of the vector $\nu^{t}$  does not depend on $t$ and it  is
subjected to the multinomial distribution:
$$
P\left\{ \nu^{t}=n_{1},\ldots , \nu_{m}^{t}=n_{m}/\varphi^{t}=0\right\} =
\left (
\begin{array}{cc}
w\\
n_{1}\ldots n_{m}
\end{array}
\right )
\prod_{j=1}^{m} P(a_{j})^{n_{j}} .
$$
That yields, along with (4),
$$
P\{\bar{\nu}^{\infty}=\bar{n}\} =P\{ \nu^{t}=\bar{n}/\varphi^{t} =0\}.
$$
It follows from this that
$$
P\{ \nu^{t}=\bar{n}\} \geq P\{\nu^{\infty}=\bar{n}\}\enskip P\{ \varphi^{t}=0\}
.$$
From this and (17) we have
$$
P\{ \bar\nu^{t}=\bar{n}\} \geq P\{\nu ^{\infty}\}\enskip P\{\varphi^{t}=0\}.
$$
That yields, along with the definition of $R^{t}$ (5),
$$
R^{t}\leq \sum_{\bar{n}\in S}P\left \{\bar{\nu}^{\infty}=\bar{n}
\right \}
\log\frac{P\{\bar{\nu}^{\infty}=\bar{n}\} }
{P\{\bar{\nu}^{\infty}=\bar{n}\} \enskip P\{\varphi^{t}=0\} }=
$$
$$
-\sum_{\bar{\nu}\in S} P\left \{
\bar{\nu}^{\infty}=\bar{n}
\right \} \log P\{ \varphi^{t}=0\}.
$$
From this and (18) we obtain   (6).

 The Theorem is proved.

   The proof of the Corollary readily follows from the  known
estimates of the number of empty boxes (see [5]).

   The proof of Theorem 3. Denote by $\pi^{t}$  the probability that
contents of some  definite  box  did  not  transform  at  the
moments $1,2,\ldots t$. Then, it is easy to see that for any box,
\begin{equation}
\pi^{t}=(1-1/w)^{t}
\end{equation}
Let us fix some letter $a_{i}\in A$  and define the random value $\Theta_{k}^{t}$
which is equal to $1$ in the case if at the moment $t$  the $k$
box contains $ a_{i}, k=0,\ldots ,w$. Then it is easy to see that $E(\Theta_{k}^{t})=
(1 -\pi^{t})\enskip P(a_{i}) +\pi^{t}\cdot E(\Theta_{k}^{0})$
and
$$
\nu_{j}^{t}=\sum_{k=1}^{j} \Theta_{k}^{t} .
$$
From the latter equalities we obtain that
\begin{equation}
E(\nu_{i}^{t})=w(1-\pi^{t})\enskip P(a_{i})+w\enskip \pi^{t} E(\Theta_{k}^{0}) .
\end{equation}
From the obvious inequality $0\leq E(\Theta_{k}^{0})\leq 1$ and from (20) we have
$$
w(1 -\pi^{t})\enskip P(a_{i})\leq E(\nu_{i}^{t})\leq w(1-\pi^{t}) P(a_{i})+
w\enskip \pi^{t} .
$$
From this, we obtain
$$
-w\enskip \pi^{t}\leq E(\bar{\nu}_{i}^{t}) - P(a_{i})w \leq w\enskip\pi^{t}.
$$
That yields, along with (19) and  the known inequality
$(1-\varepsilon ) < e^{-\varepsilon}$   the conclusion of the Theorem 3.
\newpage

\begin{center}
\Large {\bf References}
\end{center}

\begin{enumerate}
\item {\it Aho A.V., Hopcroft J.E., Ullman  J.D.}  Data Structures and
Algorithms, Addison-Wesley, Rading, MA, 1983.

\item
{\it Bell T.C., Cleary J.G., Witten  I.N.} Text  compression,
Prentice Hall, Inc., 1990.

\item
{\it Bently J.L.,   Sleator  D.D.,  Tarjan  R.E.,  Wei V.K.}
A Locally Adaptive Data Compression Scheme Comm. ACM, v.29,
1986, pp.320-330.

\item
{\it Elias P.} Interval and  Recency  Rank  Source  Coding:  Two
On-Line Adaptive Variable-Length Schemes, IEEE Trans. Inform.
Theory, v.33, N 1,1987,  pp.3-10.

\item
{\it Feller W.} An Introduction to Probability Theory  and  Its
Applications, Jon Wiley Sons, New York, vol.1, second edition,
1970.

\item
{\it Kendall  M.G.,  Stuart  A.}  The   Advanced   Theory   of
Statistics,  v.2.  (Inference  and   Relationship),   Charles
Griffin and Co.Limited, 1966.

\item
{\it Knuth D.E.} Dynamic Huffman coding,  J.Algorithms,  v.6,
1985, pp.163-180.

\item
{\it Krichevsky R.} Universal Compression and Retrieval. Kluwer
Academic Publishers, 1994.

\item
{\it Murrey  Sherk.}  Self-Adjusting  k-ary   Search   Tress,
J.Algorithms, v.19, N 1, 1995, pp.25-44.

\item
{\it Rissanen J.} Complexity of Strings in the Class of Markov
Sources".  IEEE  Trans.  Inform.  Theory, v.32, N  4,  1986,
pp.526-532.

\item
{\it Ryabko B.Ya.} Information Compression by  a  Book  Stack,
Problems  of  Information  Transmission,  v.16,  N  4,  1980,
pp.16-21, (in Russian).

\item
{\it Ryabko B.Ya.} A locally adaptive  data  compression  scheme
(Letter), Comm. ACM, v.30, N 9, 1987, p.792.

\item
{\it Ryabko B.Ya.} A Fast On-Line Adaptive Code, IEEE  Trans.
Inform. Theory, v.38, N 4, 1992, pp.1400-1404.

\item
{\it Ryabko B.Ya.} The complexity and Effetiveness of Prediction
Algorithms, J. of Complexity, v. 10, 1994, pp. 281-295.

\item
{\it Viller J.S.,  Krishnan  P.} Optimal  Prefetching  via  Data
Compression  //  Journal  of  the  ACM,  v.43,  N  5,   1996,
pp.771-793.

\item
{\it Ziv J., Lempel A.} A univercal  algorithm  for  sequential
data compression, IEEE Trans. Inform.  Theory,  v.23,  N  3,
1977, pp.337-343.

\end{enumerate}
\end{document}